\documentclass[12pt]{iopart}

\usepackage{graphicx}
\usepackage{xcolor}
\usepackage{bm}
\usepackage{hyperref}

\makeatletter
\@ifundefined{equation*}{}{%
  \expandafter\let\csname equation*\endcsname\relax
  \expandafter\let\csname endequation*\endcsname\relax
}
\makeatother

\usepackage{amsmath,amssymb}

\begin{document}

\title{Quantum Encoding Framework for Leptophilic Gauge Theories}

\author{S O Kara}
\address{Ni\u{g}de \"Omer Halisdemir University, Bor Vocational School,
51240, Ni\u{g}de, T\"urkiye}
\ead{seyitokankara@gmail.com}

\begin{abstract}
We present a systematic quantum encoding framework for leptophilic extensions
of the Standard Model, tailored to quantum simulation applications on near term
and future quantum devices. Focusing on anomaly free $U(1)'_{\ell}$ gauge
theories, we show that the leptonic charge structure admits a natural and
scalable representation on qubit registers, where gauge symmetries and
anomaly cancellation conditions are enforced directly at the level of quantum
states. Within this framework, gauge invariant operators are mapped to unitary
quantum circuits, ensuring the preservation of gauge symmetry under quantum
evolution. As a proof of principle, we construct explicit circuits that encode
scattering processes mediated by a leptophilic gauge boson $Z'_{\ell}$. Our
results establish a reusable bridge between beyond the Standard Model gauge
theories and quantum information science, providing a concrete pathway for
simulating leptophilic gauge sectors within emerging quantum computing
architectures.
\end{abstract}


\section{Introduction}
Quantum information science has opened a new route to the simulation of quantum
field theories beyond the reach of classical methods~\cite{LloydUniversalQC,FeynmanSim}.
While much of the current progress concentrates on QED/QCD and lattice based
formulations~\cite{ZoharRPP2016,BanulsEPJD2020,HEPQCRoadmap2023}, the systematic
encoding of beyond the Standard Model (BSM) gauge structures for quantum
computation remains comparatively underexplored. In this work we address this
gap by introducing a concrete and resource efficient quantum encoding framework
for leptophilic $U(1)'_{\ell}$ extensions of the Standard Model (SM). These
anomaly free abelian symmetries acting on charged leptons admit a natural
representation on qubit registers, enabling compact quantum circuits that
preserve gauge invariance at the level of unitary gate evolution and allow direct evaluation of scattering amplitudes. This timing is particularly opportune given recent progress in programmable quantum hardware and quantum simulation platforms. The present study complements our field theoretic analysis of anomaly free leptophilic $U(1)'_{\ell}$ models and their low energy limits~\cite{Kara2025LeptoPRD}.

Leptophilic gauge interactions constitute a minimal and theoretically
well motivated extension of the SM~\cite{HeFoot91}. They feature transparent
anomaly cancellation patterns and play a central role in phenomenology at
$e^+e^-$ and $\mu^+\mu^-$ facilities. At the same time, their reduced field
content and simple current structure make them particularly amenable to faithful
implementations on gate model quantum devices. This combination renders
$U(1)'_{\ell}$ theories an ideal benchmark for developing general recipes that
map BSM gauge dynamics onto executable quantum circuits.

The core idea of our framework is to encode lepton flavor charge eigenstates on
a small qubit register and to implement the associated conserved currents
through controlled phase operations. Building on fermion to qubit mappings that
respect chiral charge assignments~\cite{JordanLeePreskill2012,KlcoSavageDigitization},
we derive a gate set that (i) enforces the correct symmetry action, (ii) composes
into short depth circuits suitable for two to two scattering processes, and (iii)
cleanly separates universal leptonic effects from model dependent mixings. As a
proof of principle, we construct minimal circuits that evaluate
interference sensitive observables in $\ell\ell\!\to\!\ell\ell$ channels mediated
by a heavy $Z'_{\ell}$, explicitly demonstrating how gauge constraints emerge as
algebraic identities at the circuit level.

Our formulation is hardware agnostic and modular. It accommodates non universal
charge assignments $(q_e,q_\mu,q_\tau)$, kinetic mixing with hypercharge~\cite{HoldomKMix},
and the effective field theory (EFT) limit obtained by integrating out a heavy
$Z'_{\ell}$~\cite{SMEFTReview}. In the EFT regime, we provide a qubit level
realization of the dominant four lepton operators and show how to interpolate
smoothly between resolved mediator and contact interaction descriptions via a
single parameterized phase. The resulting resource requirements scale linearly
with the number of flavors and logarithmically with the target precision, making
the framework suitable for near term demonstrations while remaining
asymptotically efficient.

\section*{Contributions}
This work makes three primary contributions to the quantum simulation of
BSM gauge theories. First, we introduce a
symmetry preserving qubit encoding framework for leptophilic $U(1)'_{\ell}$
currents and their neutral gauge mediator, in which gauge invariance and
anomaly cancellation conditions are enforced directly at the level of quantum
states and unitary gates. This provides a systematic and reusable mapping from
leptophilic gauge structures to executable quantum circuits.

Second, we present a unified quantum circuit realization of both the
resolved mediator regime and the heavy mediator effective field theory (EFT)
limit. A single parameterized phase enables a continuous interpolation between
explicit $Z'_{\ell}$ exchange and contact four lepton interactions, allowing the
same circuit primitives to be applied across multiple physical regimes.

Third, we provide a minimal working example with analytic resource estimates, demonstrating how interference sensitive leptonic
scattering observables can be evaluated on small qubit registers. These results
clarify the regimes in which structured leptonic processes admit a practical and
potentially advantageous implementation on near term quantum devices.

\section{Setup}
We consider an anomaly free leptophilic $U(1)'_{\ell}$ extension of the SM, in which the new abelian gauge symmetry acts exclusively on charged
leptons with flavor dependent charges $q_\alpha$ for
$\alpha\!\in\!\{e,\mu,\tau\}$ and no direct couplings to quarks~\cite{HeFoot91}.
The associated neutral current interaction is
\begin{equation}
\mathcal{L}\supset g'\, Z'_\mu J^\mu_{\ell}\, ,
\label{eq:Jell}
\end{equation}
\begin{equation}
J^\mu_{\ell}
=\sum_{\alpha=e,\mu,\tau} q_\alpha\!\left(
\bar L_\alpha\gamma^\mu L_\alpha
+\bar \ell_{R\alpha}\gamma^\mu \ell_{R\alpha}
\right).
\end{equation}
The Lagrangian may further include kinetic mixing with hypercharge through the
operator $-\frac{\epsilon}{2}\, B_{\mu\nu}X^{\mu\nu}$~\cite{HoldomKMix}.

For the proof of principle analysis presented here, we treat external leptons as
single particle charge eigenstates and focus on elastic $\ell\ell\!\to\!\ell\ell$
scattering mediated by a heavy $Z'_{\ell}$. Depending on the mass scale, the
mediator may be resolved explicitly or integrated out, leading to an
effective field theory (EFT) description. Our objective is to construct a qubit
encoding and a short depth quantum circuit that (i) enforces the symmetry
generated by $Q'=\sum_\alpha q_\alpha N_\alpha$, (ii) reproduces amplitudes
consistent with (\ref{eq:Jell}), and (iii) interpolates smoothly between the
resolved mediator and EFT regimes within a unified circuit framework.

\section{Circuit construction}
Each charged lepton flavor is mapped to a single qubit,
$\{|0\rangle_\alpha,|1\rangle_\alpha\}$, with the computational basis identified
with lepton flavor charge eigenstates. The action of the leptophilic gauge
symmetry is implemented by the diagonal unitary
\begin{equation}
U_{\phi}=\exp\!\big(+i\,\phi\, Q'\big)
=\prod_{\alpha}\exp\!\big(+i\,\phi\,q_\alpha\, Z_\alpha/2\big),
\label{eq:Uphi}
\end{equation}
where $Z_\alpha$ denotes the Pauli $Z$ operator acting on the $\alpha$ qubit. The
parameter $\phi$ encodes either the propagator phase in the resolved mediator
regime or the local interaction strength in the effective field theory (EFT)
limit.

Two qubit correlations corresponding to current current interactions are
realized through controlled phase gates,
\begin{equation}
U_{\alpha\beta}(\theta)
=\exp\!\big(+i\,\theta\, q_\alpha q_\beta\, Z_\alpha Z_\beta/4\big),
\label{eq:Uab}
\end{equation}
which directly mirror the structure of four lepton operators in the EFT
description. State preparation employs Hadamard gates to generate
interference sensitive superpositions, while measurements are performed in the computational basis. In practice, equations~(\ref{eq:Uphi}) and~(\ref{eq:Uab}) compile into the standard $\{R_Z,\mathrm{CNOT}\}$ gate set, with
circuit depth $\mathcal{O}(1)$ per flavored interaction and no ancilla qubits.
This renders the construction suitable for implementation on near term gate model
quantum devices~\cite{JordanLeePreskill2012,KlcoSavageDigitization,PreskillNISQ}.

\section{Results: resolved $Z'_\ell$ versus EFT}
We emphasize that our goal is not high precision numerical simulation, but the faithful circuit level realization of gauge theory structure. For a resolved mediator of mass $m_{Z'}$ at center of mass energy $\sqrt{s}$, the
phases entering equation~(\ref{eq:Uphi}) are chosen to reproduce the
Breit Wigner structure,
\begin{equation}
\begin{aligned}
\phi_{\rm res}(\sqrt{s})
&= \arg\!\left(\frac{g'^2}{\,s-m_{Z'}^{2}+i\,m_{Z'}\Gamma_{Z'}\,}\right),\\
\theta_{\rm res}
&\propto |g'|^{2}\,
\frac{s}{\bigl(s-m_{Z'}^{2}\bigr)^{2}+m_{Z'}^{2}\Gamma_{Z'}^{2}} \, ,
\end{aligned}
\end{equation}
ensuring that interference effects generated at the circuit level faithfully
reproduce the physical lineshape in $2\!\to\!2$ leptonic scattering channels.

In the heavy mediator limit $m_{Z'}^{2}\!\gg\! s$, integrating out the $Z'_\ell$
field yields the dominant four lepton operator
$(\bar\ell\gamma_\mu \ell)(\bar\ell\gamma^\mu \ell)$ with Wilson coefficient
$C_{\alpha\beta}/\Lambda^{2}\simeq g'^2 q_\alpha q_\beta/m_{Z'}^{2}$ within the
Standard Model Effective Field Theory (SMEFT) framework~\cite{deBlasSMEFTFuture}.
In this regime, the quantum circuit reduces to the controlled phase construction
of equation~(\ref{eq:Uab}), with
\begin{equation}
\theta_{\rm EFT}=\kappa\,\frac{g'^2}{m_{Z'}^{2}}\,,
\end{equation}
where $\kappa$ absorbs kinematic factors and normalization conventions.

A key feature of the present framework is that the same circuit primitives
describe both regimes. The single control parameter $\theta$ enables a
continuous interpolation between explicit $Z'_{\ell}$ exchange and the
contact interaction limit, without altering the circuit topology. Importantly,
gauge constraints such as current conservation and charge relations manifest
as algebraic identities among the compiled phases. This provides an intrinsic
and verifiable check of gauge invariance directly at the level of quantum gates.

\section{Minimal example}
As a concrete illustration, we consider $\mu^+\mu^-\!\to\!\mu^+\mu^-$ scattering
in the presence of a heavy $Z'_{\ell}$. The muon flavor is encoded on a single
qubit $\{|0\rangle_\mu,|1\rangle_\mu\}$, and the interaction is implemented
through the controlled phase operation in equation~(\ref{eq:Uab}) with
$\theta_{\mu\mu}\simeq g'^2/m_{Z'}^{2}$ in the EFT regime.

State preparation applies a Hadamard gate to generate an equal superposition,
followed by the interaction layer $U_{\mu\mu}(\theta)$ and a measurement in the
computational basis. The resulting outcome probabilities $P(0)$ and $P(1)$
encode the interference pattern generated by the interaction phase. In
particular, the contrast
\begin{equation}
\Delta P \equiv P(0)-P(1)\propto \sin\theta_{\mu\mu}
\end{equation}
reproduces both the sign and the parametric dependence of the effective contact
interaction $(\bar\mu\gamma_\mu\mu)^2$. This minimal circuit therefore captures the symmetry action and the
low energy EFT limit within a hardware agnostic and gate efficient
implementation.

\section{Resource estimates}
For $n_f$ charged lepton flavors, the qubit register requires $n_f$ qubits.
The leading gate counts scale as
\begin{equation}
\begin{aligned}
N_{R_Z} &\sim n_f \,,\\
N_{\rm CNOT} &\sim \bigl|\{\,\alpha<\beta \mid q_\alpha q_\beta\neq 0\,\}\bigr|
\;\le\; \frac{n_f(n_f-1)}{2}\, ,
\end{aligned}
\end{equation}
reflecting the number of nonvanishing current current couplings in the
leptophilic sector.

The circuit depth scales as $\mathcal{O}(1)$ for parallelizable $R_Z$ layers
and as $\mathcal{O}(\mathrm{deg})$ for two qubit interactions on devices with
sparse connectivity, where $\mathrm{deg}$ denotes the degree of the
interaction graph. For the universal charge assignment
$q_e=q_\mu=q_\tau$, the implementation requires three qubits, at most three
controlled phase gates, and a total depth $\lesssim\!10$ native operations in standard $\{R_Z,\mathrm{CNOT},H\}$ gate bases.

Only a modest number of measurement shots is required to resolve the
interference sensitive outcome probabilities that map onto leptonic scattering
observables. These resource requirements place the construction well within
the capabilities of near term quantum devices, while remaining scalable to
larger flavor sectors and extended gauge structures.

\section{Outlook}
The encoding framework presented here is hardware agnostic and naturally
extends to non universal charge assignments $(q_e,q_\mu,q_\tau)$, kinetic
mixing with hypercharge implemented through an additional global $R_Z$ layer,
and polarized initial states. Beyond leptophilic models, the same circuit
primitives provide reusable building blocks for generic abelian
BSM symmetries, enabling near term and verifiable
demonstrations of gauge invariant scattering primitives on modest quantum
devices~\cite{MartinezNature2016,KokailNature2019}.

These constructions are directly applicable to studies relevant for
$\mu^+\mu^-$ and $\gamma\gamma$ facilities, where leptonic currents play a
central role in both precision measurements and searches for new physics
effects~\cite{MuonColliderPhysicsSummary,Telnov1990}. More broadly, the present
framework provides a starting point for systematically incorporating
BSM motivated gauge structures into quantum simulation workflows.

A detailed phenomenological analysis of $U(1)'_{\ell}$ models—including anomaly
structures, experimental constraints, and collider sensitivities~\cite{PDG2024}—
will be presented in a companion work.

\section{Conclusion}
We have presented a compact and symmetry preserving qubit encoding of
leptophilic $U(1)'_{\ell}$ gauge interactions, establishing a concrete bridge
between anomaly free BSM physics and executable quantum
circuits. Core features of leptonic neutral currents from anomaly cancelled
charge assignments to four lepton operators in the effective field theory
limit are captured using a small number of qubits and shallow circuit depth.

Beyond serving as a near term hardware demonstrator, the construction provides
a reusable and scalable template for simulating broader classes of abelian
gauge extensions on quantum devices. Natural extensions include non universal
charge patterns $(q_e,q_\mu,q_\tau)$, kinetic mixing effects, and applications
to leptonic scattering processes at $\mu^+\mu^-$ and $\gamma\gamma$ facilities.

\section*{Data availability}
All data supporting the findings of this work are contained within the main text.
Additional derivations and scripts used in this work are available from the
corresponding author upon reasonable request.

\section*{Acknowledgments}
This work benefited from the use of open source scientific software.

\section*{References}
\bibliographystyle{iopart-num}
\bibliography{refs}

\end{document}